\documentclass[12pt,a4paper]{article}
\usepackage{amsmath}
\usepackage[dvips]{graphicx}
\input epsf
\usepackage{times}
\begin{document}
\begin{titlepage}
\title{Energy evolution of the large-$t$ elastic
       scattering and its correlation with
               multiparticle production}
\author{S.M. Troshin\\
\small  \it Institute for High Energy Physics,\\
\small  \it Protvino, Moscow Region, 142280, Russia} \normalsize
\date{}
\maketitle

\begin{abstract}
It is emphasized that the collective dynamics  associated with color confinement is dominating  
over    a point-like mechanism related to a scattering of the proton constituents at the currently available values of the
momentum transferred in proton elastic scattering at the LHC. Deep--elastic scattering  and its role in the  dissimilation  
of the absorptive and reflective asymptotic scattering mechanisms  are discussed with emphasis on
the experimental signatures associated with the multiparticle production processes.
\end{abstract}
\end{titlepage}
\setcounter{page}{2}

\section*{Introduction}
Studies of elastic hadron scattering  where  initial particles are keeping their identity  can lead to a new
knowledge on  the nonperturbative dynamics of hadronic interactions, mechanism of confinement 
and asymptotic regime of strong interactions.  

Concerning relation to the color confinement phenomena,  it should be noted that according to 
the superselection rules (SSR) colored quarks and gluons live in the coherent Hilbert subspaces, and those
 are different from the physical Hilbert subspace populated by the white hadrons. 
No self-adjont operator (\it{related to the observable quantity}\rm) describing transition between colored and
bleached Hilbert subspaces can exist. It means that the color degrees of freedom can never be observed. It is the result of
SSR for the color degrees of freedom which is  combined with the non-abelian nature of QCD  \cite{tanimura}.  
But it is not a proof of confinement yet --- according to it,
color should be confined inside hadron. 
There is no known dynamical mechanism providing this nowadays.  Indeed, 
what is  known is that such mechanism should be based on the collective dynamics of quarks and gluons and,
as it was demonstrated in \cite{confus}, the unitarity might be a consequence of the confinement.

We discuss here possible manifestations of the collective effects in hadron elastic scattering keeping in mind the 
 connection of these effects with phenomena of color confinement.

\section{Coherence in the elastic scattering}
In this Section  the large-$t$ elastic scattering discussed. It should be noted that in the region of the transferred momenta beyond the second maximum in 
the differential cross-sections,  additional
dips and bumps are absent.  This smooth decrease can be considered as a manifestation of the composite  hadron structure,
then a power-like dependence can be used as a relevant function reproducing the experimental data behavior. 
On the other hand,  the exponential  function can also be applied. Those dependencies are based on the different dynamical mechanisms, namely, 
power--like behavior corresponds to the composite scattering dynamics where coherence is absent and point--like 
constituents being independent, while the exponential form should be
associated with  coherent collective interactions.

The power-like parametrization 
$d\sigma/dt \sim |t|^{-7.8}$
has been applied for the  description of the differential cross-section in the region between 1.5 (GeV)$^2$ and 2.0  (GeV)$^2$ 
in the  paper \cite{totem}. 
This dependence is depicted on the Fig. 1 (red line).
\begin{figure}[hbt]\label{78}
\begin{center}
\resizebox{6cm}{!}{\includegraphics*{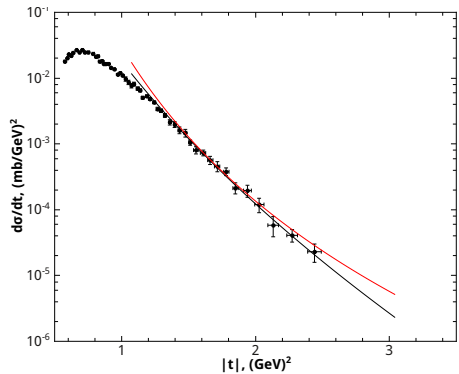}}
\end{center}
\caption{{\small \it Dependence of the large-$t$ elastic scattering differential cross-section.\rm}}
\end{figure}
At the LHC energy $\sqrt{s}=7$ TeV the power-like dependence allows to fit data in the rather narrow region of the transferred momenta.
At the same time the Orear dependence of the form \cite{orear} 
\begin{equation}\label{or}
d\sigma/dt \sim\exp(-c_{\mbox{o}}\sqrt{-t}) 
\end{equation}
can describe  the experimental data  better with $c_{\mbox{o}}\simeq 12$  (GeV)$^{-1}$, cf. Fig. 1 (solid line). 
The slope parameter  is about twice as much bigger compared to the  value of  
$c_{\mbox{o}}$ at the CERN ISR and at lower energies \cite{hart}. 
It is evident that the exponential dependence on $\sqrt{-t}$ 
 describes  experimental data in the wider region of $-t$-values and use of the power-like dependence
for the data analysis seems to  be premature and  misleading.

Different various dynamical mechanisms can provide  the Orear dependence and all of them are associted with the non-perturbative dynamics
of white hadron interactions. 
For the first time such dependence has been obtained in the multiperipheral model \cite{amati}, it has also been interpreted as a result 
of scattering into a classically
prohibited region in \cite{allil} and as one originating from the contribution of the branching point in the complex angular momentum plane in 
\cite{anselm}\footnote{I am indebted to S.S. Gershtein and L.N. Lipatov for bringing the papers \cite{allil} and \cite{anselm}, respectively, to my attention.}.
The presence of poles in the complex impact parameter plane which can result \cite{ech} 
from the rational form of the scattering amplitude unitarization leads to such dependence of the scattering amplitude too. 
For the case of pure imaginary scattering amplitude the poles in the impact parameter plane provide the additional oscillating factors in front
of the Orear exponent in the amplitude.  Such oscillations are common for the picture of diffractive scattering.
The absence of the oscillations at lower energies in the region of large $-t$    can be explained 
 by the significant role of the phase but this explanation could stop working at the LHC energies. 

Alternatively, the smooth dependence of the differential cross--section observed at lower energies can  be associated with 
 the presence of the essential double helicity-flip amplitude contribution. 
It has been shown that  the double helicity--flip amplitudes $F_2$ and  $F_4$ are important at large values of $-t$
and compensates oscillations of the helicity non-flip amplitudes \cite{dbf}. 

If the spin effects can be neglected at the LHC energies, i.e. any helicity-flip 
amplitudes would not survive at such high energies, it would result  in appearance
of the  oscillations at higher $-t$-values. 
Thus, a possible appearance of the above oscillations in the differential cross-section at higher values of $-t$
can be interpreted in this case as an observation of the $s$-channel helicity conservation 
in $pp$-scattering at the LHC energies.
\section{Asymptotics: reflective vs black disk}
The existing experimental accelerator and cosmic rays data set for the total, elastic and total inelastic cross--sections
cannot lead to the definite conclusion on the possible asymptotic hadron scattering mechanism. 
Therefore one should try to search for the independent experimental manifestations of the possible asymptotic mechanism.
In this connection it is instrumental to consider a deep--elastic scattering.  
The notion of deep--elastic  scattering introduced in the paper \cite{is} uses an analogy with the
deep-inelastic scattering and refers to the elastic scattering with the large transferred momenta $-t>4$ (GeV/c)$^2$.

With the elastic scattering amplitude being a purely imaginary function, ($f\to if$),
the function $S(s,b)$ becomes real ($S=1+2if$) and can be interpreted as a survival amplitude of the prompt elastic channel.
The relevant expressions for the survival amplitude $S(s,b)$ are the following
\begin{equation}\label{r1}
 S(s,b)=\pm \sqrt{1-4h_{inel}(s,b)},
\end{equation}
i.e. the probability of absorptive (destructive) collisions is $1-S^2(s,b)=4h_{inel}(s,b)$ ($h_{inel}(s,b)\leq 1/4$).
Simultaneous vanishing of elastic and inelastic scattering amplitudes at $b\to\infty$ should always take place and therefore
only one root in Eq. (\ref{r1})  (with plus sign) being usually taken into account, while another one (with minus sign) is omitted as a rule. This is
a well known shadow approach to elastic scattering. This is only valid   in the case when $h_{inel}(s,b)$ is
a monotonically decreasing function of the impact parameter and reaches its maximum value at $b=0$.
Thus, the inelastic overlap function has a central impact parameter profile and approaches its maximum value $1/4$,
i.e. $h_{inel}(s,b=0)\to 1/4$ at $s\to\infty$. The survival amplitude described above,  vanishes in the high energy limit
in central hadron collisions, $S(s,b=0)\to 0$.  However, the 
self-damping of inelastic channels at very high energies would lead
to a peripheral dependence on the impact parameter of the  inelastic overlap function $h_{inel}(s,b)$, it is   vanishing  at $b=0$ in the
high energy limit $s\to\infty$.  In this limit the  inelastic overlap function $h_{inel}(s,b)$  reaches its maximum value at nonzero values 
of impact parameter $b=R(s)$ \cite{ij}.
This conclusion results from the unitarity saturation by the elastic amplitude when
$f(s,b)\to 1$ at $s\to\infty$ and $b=0$.  This saturation can be realized in the framework of the rational form of unitarization (cf. e.g. \cite{ij}). 
Thus, we should take $S(s,b)=-\sqrt{1-4h_{inel}(s,b)}$
when $1/2<f(s,b)<1$.  The scattering dynamics starts to be reflective in the region where very
 high energies combined with small and moderate values of $b$  (it means that hard core appears) and approaches
asymptotically  to the completely
 reflecting limit ($S=-1$) at $b=0$ and $s\to\infty$ since $h_{inel}(s,b=0)\to 0$.
The probability of reflective scattering at $b<R(s)$  is to be determined then by the magnitude
 of $S^2(s,b)$. 

Thus, the deep--elastic scattering (DES) is associated with reflective scattering at very high energies where the
colliding hadrons  do not suffer from absorption anymore.
The DES dominates over multiparticle production (at small impact parameter values
 $h_{inel}(s,b=0)\to 0$).  This ensures favorable conditions for the experimental measurements, since the peripheral
 profile of $h_{inel}(s,b)$,  associated with reflective scattering,
suppresses the probability of the inelastic collisions in the region of small
impact parameters. The main contribution to the mean multiplicity is due to
the peripheral region of $b\sim R(s)$. DES in this case is correlated with inelastic events of low cross--sections,
i.e. it has a small background due to production and high experimental visibility. The reflective  mechanism  associated with the
complete unitarity saturation will asymptotically decouple from particle production asymptotically and at finite energies
 it corresponds to observation of the DES
 with   decreasing correlations  with   particle production. Contrary, the saturation of the black disk limit
implies strong correlation of DES with multiparticle production processes \cite{plbus}.

\section*{Acknowledgement} I am pleased to express the gratitude
to IHEP, Protvino and Organizers of Diffraction 2012 in Puerto del Carmen, Lanzarote, 
for the support of my participation to the Conference. The talk is based in part on the papers prepared in collaboration with N.E. Tyurin. I am grateful to 
him for 
the many useful comments and discussions. I would also like to thank L.L. Jenkovszky, U. Maor, V.A. Petrov and N.P.~Zotov for the interesting remarks
and discussions.

{\small 

\end{document}